\documentclass[11pt]{article}

\input psfig
\def\beq{\begin{eqnarray}}
\def\eeq{\end{eqnarray}}
\def\bea{\begin{eqnarray*}}
\def\eea{\end{eqnarray*}}

\def\centeron#1#2{{\setbox0=\hbox{#1}\setbox1=\hbox{#2}\ifdim
\wd1>\wd0\kern.5\wd1\kern-.5\wd0\fi
\copy0\kern-.5\wd0\kern-.5\wd1\copy1\ifdim\wd0>\wd1
\kern.5\wd0\kern-.5\wd1\fi}}
\def\ltap{\;\centeron{\raise.35ex\hbox{$<$}}{\lower.65ex\hbox{$\sim$}}\;}
\def\gtap{\;\centeron{\raise.35ex\hbox{$>$}}{\lower.65ex\hbox{$\sim$}}\;}

\def\singleandthirdspaced{\baselineskip=\normalbaselineskip\multiply
    \baselineskip by 130\divide\baselineskip by 100}
\def\singlespaced{\baselineskip=\normalbaselineskip}

\newcommand{\newc}{\newcommand}
\newc{\qbar}{{\overline q}}
\newc{\Kahler}{K\"ahler }
\newc{\deltaGS}{\delta_{\rm GS}}
\begin{document}
\begin{titlepage}
\begin{flushright}
{\large hep-th/9906246 \\ SCIPP-99/23\\ PUPT-1875

}
\end{flushright}

\vskip 1.2cm

\begin{center}

{\LARGE\bf Remarks on the Racetrack Scheme}

\vskip 1.4cm

{\large Michael Dine$^a$ and Yuri Shirman$^b$}
\\
\vskip 0.4cm
%{\it $^a$Stanford Linear Accelerator Center,
%     Stanford CA 94309} \\
{\it $^a$Santa Cruz Institute for Particle Physics,
     Santa Cruz CA 95064  } \\
{\it $^b$Physics Department, Princeton University, Princeton, NJ
08544 }
\\

\vskip 4pt

\vskip 1.5cm

\begin{abstract}

There are only a small number of ideas for stabilizing the moduli
of string theory.  One of the most appealing of these is the
racetrack mechanism, in which a delicate interplay between two
strongly interacting gauge groups fixes the value of the coupling
constant.   In this note, we explore this scenario. We find that
quite generally, some number of discrete tunings are required in
order that the mechanism yield a small gauge coupling.  Even then,
there is, in general, no systematic weak coupling approximation.
On the other hand, certain holomorphic quantities can be computed,
so such a scheme is in principle predictive. Searching for models
which realize this mechanism is thus of great interest.  We also
remark on cosmology in these schemes.
\end{abstract}

\end{center}
 \vskip 1.0 cm

\end{titlepage}
\setcounter{footnote}{0} \setcounter{page}{2}
\setcounter{section}{0} \setcounter{subsection}{0}
\setcounter{subsubsection}{0}

%%%%%%%%%%%%%%%%%%%%%%%%%%%%%%%%%%%%%%%%%%%
%%%%%%%%%%%%%%%%%%%%%%%%%%%%
\singleandthirdspaced

%\begin{document}

\section{Introduction}

Understanding how the moduli of string or $M$-theory
compactifications are fixed is one of the greatest challenges of
the subject.  For compactifications with more than four
supersymmetries ($N>1$ in four dimensions), general considerations
suggest that there is an exact moduli space. In the case of four
or less supersymmetries, generically the flat directions are
lifted, with the potential typically tending to zero in any region
in which the appropriate couplings tend to zero or the radii tend
to $\infty$.\footnote{In compactifications of non-supersymmetric
string theories, the story is slightly different.  Generically, in
the loop expansion, one finds a non-zero cosmological constant, as
a result of which the energy may blow up for large radius, but
will still tend to zero for small coupling.  An example of an
$N=1$ theory where some of the moduli are exact was given
in~\cite{bdmoduli}.}

This argument suggests that the string coupling should be strong,
and the scales of the theory should be
comparable~\cite{dineseiberg}.   This in turn raises the question:
why are the observed gauge couplings in nature weak and unified,
and why does the unification scale seem to differ significantly
from the Planck scale?   This is the real issue in stabilization
of the moduli:  given that stabilization must occur, if at all, in
regions where no sort of weak coupling approximation
can be valid, why should anything be
calculable?  It is, after all, not hard to imagine schemes to
stabilize the moduli, but it is hard to see why the coupling
should be small, except as a result of numerical accidents,
shrouded in mysterious high energy physics.  In this view, none of
the parameters of low energy physics would be calculable in any
systematic approximation scheme.

These points are illustrated by the various toy models of modulus
stabilization which appear in the literature. Most of these focus
on a single modulus, and postulate superpotentials from one or
another source which provide stabilization~\cite{dudas,nilles,dvali}.
Generically, however, the couplings turn out to be of order one in
these proposals, and it is necessary to suppose that uncontrolled
strong coupling dynamics explains why the coupling is small. These
models certainly illustrate that the gauge coupling can be small,
but they do not predict that any of the parameters of
low energy physics should be calculable. It is perhaps worth
noting that one {\it can} construct weakly coupled string models
with (at tree level), as few as a single modulus~\cite{dh}, or in
which all moduli or all moduli but one are charged under discrete
symmetries.   One can also contemplate theories with no moduli at
all~\cite{dinesilverstein} or in which all moduli transform under
unbroken symmetries, but it is unclear whether any controlled
approximation might be available.

There are only a small number of proposals for fixing moduli in
which some quantities are calculable. One is known as ``\Kahler
stabilization"~\cite{coping}. Here one imagines starting with some
weakly coupled limit of $M$ theory (i.e. some limit in which a
systematic approximation is available), where one can
calculate holomorphic quantities such as the gauge coupling
functions and the superpotential.   In other words, as one takes
some modulus to extreme values, ${\cal M}\rightarrow \infty$, one
comes to a regime where one can perform systematic calculations in
${\cal M}^{-1}$.  The superpotential and gauge couplings are
holomorphic, and because of discrete shift symmetries, they are
functions of $e^{-{\cal M}}$.   As one increases the couplings one
supposes that, in a regime where the exponential is small,  there
are large corrections to the \Kahler potentials of the moduli, such
that the potential has a minimum at weak coupling. Any holomorphic
quantity which can be computed in the weak coupling limit will be
calculable in such a picture.

A second proposal involves the possibility of ``maximally enhanced
symmetry"~\cite{dns}.  Here one argues that the minimum of the
full potential might naturally lie at a point where all of the
moduli transform under unbroken symmetries.  In addition to the
fact that such points are automatically stationary points of the
effective action, this hypothesis naturally solves the moduli
problem of string cosmology. However, generically in such states,
$\alpha \approx 1$. One must hope that there are some such
states for which the effective low energy couplings
happens to be small (and unified). Little is
calculable in such a picture; however, this hypothesis leads to
the prediction that supersymmetry will be broken at low energies,
and that some sort of gauge mediation will play a crucial role. A
third possibility is that there are simply no moduli.
Operationally, this hypothesis is similar to that of maximally
enhanced symmetry. In the context of large extra dimensions,
possibilities involving large topological charges have been
proposed~\cite{sundrum,dimopoulosmoduli,bdn}, and in some cases,
the size of the extra dimension is correlated with the smallness
of the gauge couplings~\cite{yukawa}. Finally, we will focus in
this note on the ``racetrack
mechanism"~\cite{racetrack,taylor,casas,kl}.

The racetrack proposal is in some sense more ambitious than the
others we have listed. Here one hopes for a systematic analysis of
moduli stabilization in the low energy effective field theory. The
basic idea is that competing effects from different low energy
gauge groups may give rise to a local minimum for the moduli, in a
computable fashion, at weak coupling. One might then hope to
compute other quantities relevant to low energy physics.

From the beginning, questions have been raised about calculability
in this scenario~\cite{seibergpc,bankspc}.  This question will be
a central focus of our investigation. We will see that in some
versions of the racetrack scheme, nothing is computable and that
most quantities are  not likely to be computable in {\it any}
circumstance.  But upon more careful consideration, it will become
clear that the racetrack scenario has many features in common with
the \Kahler stabilization and maximally enhanced symmetry
hypotheses in that in some cases, holomorphic quantities such as
the gauge couplings and superpotential are computable. To simplify
the discussion, we will assume unification, so the standard model
gauge couplings are controlled by a single modulus, which we will
loosely refer to as the ``dilaton."  We point out that in order to
have any control over low energy physics, it is necessary that the
scale of the gauge groups be hierarchically small, i.e. that it be
much below the fundamental scale. This requires one (discrete)
fine-tuning. Even then, one is unlikely to be able to compute the
\Kahler potential in a systematic weak coupling approximation;
there is, in general, no quantity which one can take arbitrarily
small in order to justify such a calculation.   Holomorphic
quantities, however, may be computable, just as in the case of
\Kahler stabilization. In other words, one can compute holomorphic
quantities
at weak coupling, and these computations should be reliable at the
true minimum.  The point is simply that, by symmetries and
holomorphy, corrections to holomorphic quantities are controlled
by powers of $e^{-S}$, so that if this quantity is hierarchically
small, corrections are similarly small.

This is not the case for the Kahler potential and other
non-holomorphic quantities.  No simple symmetry argument controls
its dependence on $S$.  In some instances, we can exhibit large
corrections to the \Kahler potential by examining the low energy
effective field theory.  If we assume that the appropriate cutoff
for this theory is that governed by the relations between
couplings and scales of the weakly coupled heterotic string,
corrections to the \Kahler potential, as we will see, are formally
of order one.  If we take the scalings suggested by the
Horava-Witten picture,the story is somewhat different.  In this
limit, corrections to the Kahler potential are suppressed by
powers of $\rho$, the size of the eleventh dimension.  Corrections
to higher derivative operators in the low energy field theory are
of order one.  If one assumes that a Type I picture is valid, as
we will see, corrections to quantities in the low energy theory
are suppressed by ${g_{eff}^2\over 8 \pi^2}$.

Given that one cannot make the gauge groups (and hence the
coupling constants) arbitrarily small in these limits, the message
we take from these observations is that generally only holomorphic
quantities will be calculable.   Still, we cannot rule out the
possibility that we might be lucky, and that leading order
computations might be reliable for other quantities as well.

Apart from these differences, it is perhaps worthwhile to
distinguish two cases:  supersymmetry unbroken at the minimum of
the potential, and supersymmetry broken.  In the latter case,
because of the potential problem of computing the Kahler
potential, it is difficult to perform any analysis. If one
supposes that the \Kahler potential is calculable, then one can
compute the cosmological constant; for
typical forms of the superpotential, it is
unlikely to vanish.

In the case that the potential for $S$ does not break
supersymmetry, one has, in principle, more control. Holomorphic
quantities are computable.  Moreover, as will be described
elsewhere, such a situation might be desirable for cosmology. In
this case, some other sector of the theory must be responsible for
supersymmetry breaking.  If gravity is the principle messenger of
supersymmetry breaking (as is possible if the symmetries are not
enhanced), the non-calculability of the \Kahler potential means
that one has little control over the low energy dynamics. Soft
breakings, in particular, are not computable. The situation is
potentially quite different if supersymmetry is broken by low
energy dynamics, as in gauge mediated models. The gauge couplings
and some Yukawa couplings (i.e. ratios which depend only on
holomorphic quantities) would be computable. The soft breakings
would be computable in terms of a small number of parameters (some
terms in the low energy effective superpotential would depend on
uncontrollable \Kahler potential corrections). Many of the
uncomputable non-holomorphic quantities may be of little relevance
to low energy physics.

The racetrack explains how one modulus is fixed, with a large
value for its mass.  If there are other moduli, they may
still pose problems for cosmology.  On the other hand, if
these moduli all sit at enhanced symmetry points, the cosmological
moduli problem\cite{bkn,casasmoduli} is solved, and low
energy breaking, as in the case of maximally enhanced symmetry, is
inevitable.  In such a picture, it is quite natural that
only one modulus has a large value in fundamental units, and thus
is responsible for the observed values of the gauge couplings and
their unification.  As stressed by the authors of \cite{bkn},
any low energy supersymmetry breaking scheme has a drawback:
in order to generate a term in the superpotential of the correct
order of magnitude to lead to
vanishing cosmological constant, it is necessary to postulate
some additional strong dynamics beyond that which breaks supersymmetry.
We will make some remarks on this below.  It is not
clear whether this problem is truly more severe than the extreme
fine tuning required in any case.

This picture is indeed attractive from a cosmological point
of view, as discussed in \cite{bkn}.  These authors argued that
it might be desirable to fix the mass of some moduli at scales
well above the scale of supersymmetry breaking.  Any remaining
moduli pose potential cosmological problems, unless, as
argued in \cite{drt}, they sit at enhanced symmetry points.

Before investigating these questions, we should introduce our
basic assumptions and some terminology.  Our focus is on the
question: why are the gauge couplings small and unified.  To
address this, we will assume, as stated above, that there is one
modulus whose value controls the size of the observed gauge
couplings (other moduli, with small expectation values, could also
couple). We will refer to this modulus as the dilaton, and denote
it by $\cal M$, though we won't assume that this field is to be
identified with what is usually called the dilaton in weakly
coupled string theory. Second, we need to explain, as in any such
discussion, what is meant by the term ``modulus." Clearly, since
we are discussing the problem of stabilization, we are not
supposing that there is an exact moduli space.  We have instead in
mind the possibility that there are approximate moduli, whose
masses at their minima are small compared to the fundamental
scale, and which become exact moduli in some limit.  Finally, we
are assuming throughout that there is approximate low energy
supersymmetry. Indeed, we will see that it is hard to make sense
of the racetrack scheme without it.

One must also note that there are actually several versions of
the racetrack idea.   Most are tied specifically to gaugino
condensation, but this is not necessary~\cite{yanagida}, and, as we
will see, has certain disadvantages. All involve generating a
superpotential in the low energy theory.  In most versions of the
scenario, the dynamics which fixes the moduli does not break
supersymmetry.  We will argue that this is essential if the gauge
couplings are to be calculable. In perhaps the simplest proposal,
there are several groups coupled to the dilaton with very large
$\beta$-functions~\cite{kl,yanagida}. In this case, as we will see,
the usual low energy analysis yields a small value for the gauge
coupling. However, the scale of the low energy theory is of order
the fundamental scale, and the low energy analysis is not valid.
One might have hoped that holomorphy and symmetries would allow
one to extend the range of validity of these methods. However the
modulus superpotential, even assuming the
relevance of the low energy theory, is a much more complicated
function than usually assumed. This superpotential is not
calculable, and the mass of the modulus is of order one.
Generically, the cosmological constant will be of order one,
though it is possible, in principle, for it to vanish at this
level, due to symmetries. So, in this case, little is gained
over simply assuming that the theory has a
supersymmetry-preserving minimum at some desired value of the
coupling.

More promising is a second, (discretely) fine-tuned
version, in which several groups have very similar, large
$\beta$-functions. In this case, the coupling can be small, and
there can be a hierarchy of scales. In the case where gluino
condensation is the origin of the superpotential, at least three
groups must have nearly equal, large $\beta$-functions, and
special relations must hold among threshold factors. However,
following~\cite{yanagida}, we can consider models with unbroken
$R$ symmetries (in this reference it was supposed that the
symmetries were continuous but discrete symmetries can also
accomplish the same objectives; the role of
discrete R symmetries in obtaining unbroken supersymmetry was
first stressed in \cite{bkn}).  In this case, only one fine
tuning is required, though one also needs many gauge singlet
fields and discrete symmetries.

Even in this discretely fine-tuned case, it is unlikely that some
sort of weak coupling analysis will be possible. First, there is
no small parameter (such as $1/N$) which justifies such an
approximation.  Second, if one assumes the relations of couplings
and scales as in the weakly coupled heterotic string,
while the gauge couplings are
numerically small, it is easy to exhibit loop corrections, at
least for non-holomorphic quantities, which are of order one.  As
we have already noted, the situation is different in other
string theories, but given that one cannot obtain very
large gauge groups in these limits, we view this result as
suggestive of a more general difficulty.
However, certain holomorphic quantities are
under control, and may be susceptible to analysis in the low
energy theory.  To understand this, one should imagine first
passing to the weak coupling limit.  In this limit, high energy,
effects in the superpotential and gauge coupling function go as
powers of $e^{-{\cal M}}$.  The low energy analysis yields a
coupling such that $e^{-{\cal M}}$ is extremely small. So these
corrections should be under control.  In a scheme of this sort,
one must still understand the breaking of supersymmetry and the
fixing of any other moduli.  If the breaking is at scales
intermediate between the weak and Planck scales, (as in
``supergravity" models), low energy soft breaking is not
calculable.  If breaking is at low energies, as in gauge
mediation, many of the important features of the low energy theory
may be calculable.

Different scalings hold in the Type I or strongly coupled
heterotic string limits  and it is conceivable that the
\Kahler potential is calculable.  But even if many quantities
are not calculable,
one would be left with a rather appealing
picture. Allowing one numerical coincidence (some close relation
among beta functions), one would hope to develop a complete
phenomenology starting from a weakly coupled limit.  One would
still need to understand the problem of the cosmological constant,
of course.

We turn, finally, to the possibility that supersymmetry is broken
simultaneously with fixing the moduli.  For such scenarios, we
note that, in light of our observation that one cannot compute the
\Kahler potential, very little, if anything is accessible to
analysis. The gauge couplings would not be calculable, nor would
soft breaking terms.  It is possible that, as in the case of
Kahler stabilization, some terms in the superpotential might be.
Even if one could use the lowest order \Kahler potential (as we
will see might conceivably be the case) one has another
difficulty:  the cosmological constant is calculable,
and typically non-vanishing at the
minimum.

In the following sections, we review the racetrack idea and
describe these possibilities in greater detail.  We conclude by
arguing that indeed the racetrack idea, whether ultimately
realized in nature or not, does provide a viable model for
understanding the smallness of the gauge couplings in a theory
which is inherently strongly coupled.  As an application, we
consider the implication of such a picture for scenarios with
large internal dimensions.

\section{The SUSY Conserving Version of the Racetrack Idea}

Kaplunovsky and Louis have put forth an appealing version of the
racetrack idea. They note that studies of $F$-theory
compactifications have yielded classical ground states of the
theory with enormous gauge groups. Suppose, now, that one has two
gauge groups (without matter -- these remarks all readily
generalize to cases with matter in which the strong gauge group
does not by itself break supersymmetry) with very large
$\beta$-functions, say \beq b_1 = aN~~~~~b_2 = bN, \eeq where $N$
is a large integer, and $a$ and $b$ are (rational numbers) of
order unity. Suppose the gauge couplings of both groups are
controlled by a single modulus, ${\cal M}$, i.e. the lagrangian at
low energies looks like \beq \int d^2 \theta {\cal M} (W_1^2 +
W_2^2). \eeq Then the usual arguments for gluino condensation
yield a superpotential for ${\cal M}$: \beq W=\alpha b_1e^{-{\cal
M}/b_1} -b_2 \beta e^{-{\cal M}/b_2}.\eeq  (If the couplings are
not the same at the fundamental scale, this difference can be
absorbed into the $b_i$'s.) Here $\alpha$ and $\beta$ are numbers
of order one which arise due to threshold corrections. This
superpotential has a stationary point at \beq {\cal M} = {b_1 b_2
\over b_1-b_2} \ln({\beta \over \alpha}). \eeq For large $N$, this
behaves as \beq {\cal M} \sim N \ln(\beta/\alpha).\eeq At the
stationary point, however, $W$ is non-zero,
\beq W= \alpha(b_1-b_2) \left({\beta\over
\alpha}\right)^{-{b_2 \over b_1-b_2}}
 \eeq
(when $b_1=b_2$, there is no stationary point).

This would appear to be what is needed to stabilize the dilaton at
weak coupling, and would even seem to be rather generic.  There
are  at least two difficulties, however, which appear more or less
fundamental. The first has to do with the self consistency of the
calculation, and the second to do with the problem of the
cosmological constant.  The usual analysis of gluino condensation
assumes that the scale of the low energy gauge theory, $\Lambda$,
is well below the fundamental scale.  But in this scheme,
$\Lambda$ is of order one.  As a result, it is not at all clear
why one can look at the gauge group, and not consider the full set
of massive string (M-) theory states.

 Still, sometimes holomorphy and symmetries can significantly
constrain the form of the superpotential, so we might hope that
the low energy analysis captures some of the truth.  To assess
this possibility, let us consider the effects of
non-renormalizable operators on the low energy analysis.  The
usual discussion of gluino condensation starts by noting that at
the renormalizable level, the effective theory has an $R$ symmetry
under which the modulus ${\cal M}$, transforms.  This $R$ symmetry
uniquely determines the dependence of the condensate, $\langle
\lambda \lambda \rangle$ on ${\cal M}$, \beq \langle \lambda
\lambda \rangle = z, \eeq where \beq z=e^{-{3{\cal M} \over b_o}}.
\eeq Now suppose that there are terms in the effective action just
below the string scale of the form \beq \int d^2 \theta \chi W^4 +
\zeta W^6 + \dots \eeq We can think of $\chi$ and $\zeta$, etc.,
as spurions which transform under the $R$ symmetry; $\chi$ has $R$
charge $-2$, $\zeta$ charge $-4$, etc. The same argument which
gave the leading term gives \beq \langle \lambda \lambda \rangle =
z + a\chi z^2 + b\zeta z^3 + \dots \eeq where $a$ and $b$ are
constants of order one.

So we see that the superpotential is a general function of $z$,
even in the case of one modulus.  At the stationary point, these
corrections are not suppressed.  Indeed, even with only one
modulus there may now be a stationary point with large ${\cal M}$.
It is not possible to determine the location of this stationary
point, however, without an understanding of the full string
theory.  The low energy theory is insufficient.

In order that one obtain a supersymmetric vacuum with vanishing
cosmological constant, it is necessary that both the
superpotential and its first derivative vanish.  This is not the
case for our simplified treatment of the example above; at the
stationary point of $W$, $W$ is non-zero.  It is conceivable that
there are models for which $W$ has special properties such that it
vanishes at the stationary point.  After all, the vanishing of $W$
in weakly coupled string compactifications is a consequence of
detailed features such as the Peccei-Quinn symmetry.  It is
possible that models whose dynamics preserves a discrete $R$
symmetry could naturally yield a vanishing $W$ at the stationary
point.  We will discuss this possibility below in a different
context, but the other difficulties remain.

Finally, note that in these schemes, the modulus itself is
massive, with mass of order one.  So one might as well suppose
that one is studying $M$ theory vacua (in some approximation) with
fewer moduli from the start.  Logically, there is no problem with
this idea, but it leaves unanswered the questions of why the
couplings are weak and whether anything is calculable.

\section{Improvement through Fine Tuning}

In our example above, if $b_1 \approx b_2$, one can perhaps
improve the situation. In this case, $z$ can be small. For
example, if the two quantities are within ten percent of each
other, than $z \sim 10^{-10}$.  The scales are now well separated,
and arguably the use of the low energy effective action is
self-consistent.  The situation with respect to the cosmological
constant is also somewhat better.  For suppose supersymmetry
breaking arises from some more weakly coupled gauge group with
beta function $b_3 \ll b_1$. It is possible (for example if the
leading order contribution -- in $z$ -- to the cosmological
constant vanishes) that the various terms could cancel with one
another.

In the finely tuned case, it is possible to analyze the possible
vanishing of $W$ with several gauge groups.  With two gauge groups
there are still no solutions, but with three or more gauge groups
with very similar, large $\beta$-functions, there are solutions,
for suitable values for the threshold corrections
\cite{taylor}.  One can
analyze this problem by considering a superpotential of the form
\beq W= \alpha b_1 e^{-{\cal M}/b_1} + \beta b_2 e^{-{\cal M}/b_2}
 +\gamma b_3 e^{-{\cal M}/b_3} \eeq
We want to ask whether there can be solutions of the equations
$W^{\prime}=W=0$, for some values of $\alpha$, $\beta$ and
$\gamma$.  It is a simple algebraic exercise to verify that this
is the case.  Whether the required values of the threshold
corrections actually arise is another question, but
it is not perhaps completely implausible, given that corrections
to $\alpha,\beta$ and $\gamma$
from their one loop form will be exponentially
small.

Izawa and Yanagida have proposed a variant on the racetrack scheme
which would ameliorate this difficulty~\cite{yanagida}. Field
theories with quantum moduli spaces, at the level of
non-renormalizable terms, leave unbroken $R$ symmetries.  If such
a theory appears in the low energy limit of a string theory, and
if the theory has a (discrete) R symmetry, then the superpotential
can naturally vanish at the stationary point (the fact
that $R$-symmetries can account for unbroken supersymmetry
and vanishing cosmological constant was stressed
in \cite{bkn} and is crucial to the cosmological scenario
outlined in \cite{bankscosmology}).  These authors give
an example with two (discretely tuned) low energy groups.  A large
number (of order $N^2$) singlets with suitable couplings to the
matter fields of these groups are required to achieve the desired
stabilization. Additional discrete symmetries are required in
order to obtain the desired patterns of couplings. On the other
hand, in brane constructions such large numbers of singlets might
be plausible, and elaborate discrete symmetries are familiar in
string theory.  This type of $R$-symmetry based scenario, then,
seems the most plausible.

\begin{figure}[htbp]
\centering
\centerline{\psfig{file=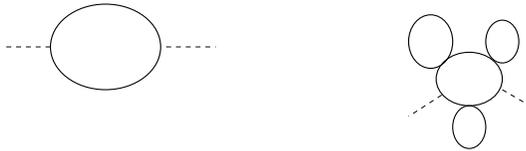,width=7cm,angle=-90}}
\caption{Contributions to dilaton \Kahler potential proportional to
$N^2$.}
\label{largeloops}
\end{figure}

Finally, there is the question of what is calculable in this
picture.  Looking at the explicit solutions, one sees that if the
$\beta$-functions are of order $N$, and their differences of order
$1$, the gauge couplings are of order $1/N^2$.  This is probably
enough to ensure that holomorphic quantities are given by their
weak coupling values.  Corrections to the superpotential and the
gauge coupling functions go as $e^{-{\cal M}}$, for example, and
this is suitably small.  However, non-holomorphic quantities are
not, in general, described by any weak coupling approximation.  In
particular, consider corrections to the \Kahler potential for
${\cal M}$, such as those indicated in fig. \ref{largeloops}.
These diagrams come with a factor of $N^2 \Lambda^2/M_p^2$, where
the $N^2$ arises from the $N^2$ particles propagating in the loop,
and $\Lambda$ is a cutoff\cite{gaillard}.  In the weakly coupled heterotic
string, $\Lambda^2 = g^2 M_p^2$, and so these corrections are of
order $1$.
The resulting one loop SUGRA contributions to the soft masses are of
the order $N^2\Lambda^2/(16 \pi^2 M_p^2)$ \cite{clm}. Taking
into account factors of $\pi$ in our definition of the
dilaton we find that they are of order $1$. We should stress
here that in some models supergravity contributions may be
numerically small.  If one examines the scenarios
discussed in ref. \cite{casas}, one finds that for most
of them, $g^2 {\sum N_i^2 \over 16 \pi^2}$
is $30\%$ or larger, but there is one where
it is as small as $10 - 15\%$.  One might hope that
the corrections can be reliably computed.
However, it will be difficult to establish
this fact, since, as we have argued, there is no formal small
expansion parameter.

So far we have focused on the weakly coupled heterotic
string theory in order to estimate the cutoff.
Some hope for optimism is provided by considering other string
theories (we are grateful to I. Antoniadis for a remark which
prompted an examination of this question).
In the limit of the strongly coupled heterotic string one
would expect the cutoff to be given by $M_{11}$. However,
for quantities which are predicted by $10$ dimensional
supersymmetry, the relevant cutoff will related to the 
compactification scale.
Support for this comes from studies of the
Kahler potential in the Horava-Witten limit, which is easily seen
by symmetry arguments to be the same as in the weak coupling
limit, up to terms of order $1/\rho$, where $\rho$ is the size of
the eleventh dimension.  This can be traced to the fact that one
can pass from one limit to the other in such a way that the theory
is always approximately ten dimensional.
So the \Kahler potential may be calculable. For
example, if the typical size of the M-theory compactification
manifold is $R > M_{11}^{-1}$, then the SUSY above the
compactification scale may provide the relevant cutoff. In
such a case corrections could be of order $N^2 R^{-2}/M_p^2
\sim N^2 g^3 \sim 1/N$. Certain higher dimensional operators would
obtain non-calculable corrections, but this could be of
little relevance to low energy effective theory.
In the Type I theory,
assuming compactification at the string scale,
$\Lambda^2 = g^4 M_p^2$, and
so again one might hope that these corrections are under control
for sufficiently large $N$ (applied to the
examples of \cite{casas}, this gives corrections in some
cases as small as $7\%$, with $20\%$ being more typical).
Of course, to argue this formally
requires that there exist a set of theories characterized by $N$
such that one can take the limit $N\rightarrow \infty$, and this
seems unlikely to exist in these limits.\footnote{The same, of course,
is true for holomorphic
quantities, but given their exponential dependence, this seems
more plausible.}  F-theory suggests that very large gauge groups
may exist, but we do not know how to perform the corresponding
analysis for $F$ theory, and suspect that one will
have similar difficulties to those of the weak coupling
heterotic picture.   We will adopt the pessimistic view in what
follows that one does not expect to be able to compute the \Kahler
potential. In this view, one does not expect to be able to
calculate quantities which depend on the detailed form of the
\Kahler potential.  But it is again important to keep in mind that
there may be instances where much more is calculable.

In sum, the finely tuned case is a scenario in which a low energy
analysis can in principle provide an explanation of small
couplings and large hierarchies.  Holomorphic quantities are in
principle calculable in such a scheme, but non-holomorphic
quantities are probably not.  If supersymmetry is broken at an
intermediate scale, it will not be possible to say much about the
low energy spectrum, since the \Kahler potential is not known.  On
the other hand, if supersymmetry is broken at low energies, as in
gauge mediation, many quantities may be calculable.  Apart from
the gauge couplings themselves, physical quantities which depend
holomorphically on terms in the superpotential will be calculable.
The soft breakings should be expressible in terms of a small
number of parameters.

\section{Supersymmetry Violating Version of the Racetrack Scheme}

When originally proposed, it was hoped that the racetrack scheme
would provide a mechanism for fixing some moduli (assumed to be
the usual dilaton of weakly-coupled heterotic string theory) while
simultaneously breaking supersymmetry in a calculable manner, and
generating a weak gauge coupling and large hierarchy. This
possibility was most thoroughly analyzed in \cite{casas}, where
many interesting examples were developed. Still, we can ask
whether such a proposal can truly be analyzed in terms of a low
energy effective action.

As in the supersymmetric case, it is necessary, in order that any
low energy effective action analysis make sense (and presumably
also that one obtain hierarchically small supersymmetry breaking)
that one has several (at least two) groups with nearly identical
$\beta$-functions.  However, determining whether a minimum of the
potential exists, and its location (and in particular determining
the value of the gauge coupling) requires, in this case, knowledge
of the \Kahler potential for the modulus. We have argued, however,
that this may not be calculable.  The problem can be stated more
strongly:  there is no simple argument, in these cases, that the
superpotential is simply a sum of the superpotentials for the
different groups.  The usual symmetry
arguments
\cite{gauginocondensation} for the form of the
superpotental no longer hold.  Moreover, at generic points in the
moduli space, the larger condensate induces non-zero -- and large
-- contributions to the other.  Examining the appropriate
diagrams, one can see that this problem is closely tied to the
problem of understanding the \Kahler potential.
\footnote{Similar issues were raised in
\cite{holomorphy}, where it was
noted that in certain grand unified theories, one obtained
inconsistencies if one assumed the superpotential was a simple sum
of this type.  In \cite{burgess}, it is asserted that the
superpotential is a sum.  However, this analysis treats the
dilaton superfield as a non-dynamical background
and ignores the fact
that one condensate induces corrections to the other.  It is
conceivable that the correct form is a sum, and that these other
effects can be absorbed into corrections to the \Kahler potential,
but this is by no means obvious, and is a
question worthy of further investigation.}

Indeed, this
situation is not so much different than that of ``\Kahler
stabilization," where it is supposed that with a single gauge
group, the \Kahler potential is such as to give rise to a minimum
of the potential at weak coupling.  The principle difference is
the two, nearly equal, $\beta$-functions provide a slightly
different explanation for the smallness of the gauge coupling than
the accident proposed in~\cite{coping}.

As we have remarked
above, the analysis of \cite{casas} is consistent with these
remarks.  In most cases, a rough estimate of the corrections
yields a large value for the effective expansion parameter,
but in one of their examples it is about 10\%.  Whether
this is good enough in string theory is, of course,
an open  question.

Operationally, it is not clear that there is much difference
between the two hypotheses.  In particular, in both cases, some
quantities protected by holomorphy, i.e. the superpotential and
gauge coupling functions, are accessible. Quantities which are
not, such as the soft breaking masses, are unpredictable.
Recently, in \cite{choi}, it has been shown that a combination of
Kahler stabilization and multiple condensates provides an
interesting model for stabilization.  A quite specific and
plausible picture for the origin of the \Kahler potential
corrections is presented, though again control of non-holomorphic
quantities, in the sense of there being a systematic, weak
coupling approximation scheme, is limited.

It should be noted that if the \Kahler potential is given by its
weak coupling (or strong heterotic coupling) form,
the cosmological constant can be calculated
at the level of the  effective action.  The dilaton
potential is (now calling the dilaton
$S$, as appropriate to the weak coupling limit),
\beq
V(S,S^{\dagger})= {1 \over S+S^{\dagger}} \left [\vert
({\partial W \over \partial S} + {1 \over S+S^{\dagger}}W\vert^2
(S+S^{\dagger})^2 -3 \vert W \vert^2 \right ].
\eeq
It is not easy to find systems whose $W$ gives minima
with $V=0$\cite{racetrack}.

\section{Other Alternatives to the Racetrack Scheme}

We have stressed in the preceding sections that the problem in
string theory is not to explain how moduli can be stabilized, but
rather how they can be stabilized in such a way that the gauge
couplings are weak and supersymmetry is hierarchically broken, and
such that anything is computable.  We have argued that the finely
tuned version of the racetrack scheme does provide a picture in
which the gauge couplings could be fixed at small values and a
hierarchy explained in a calculable fashion.  Of course, we do not
have a detailed string model which realizes these ideas, but at
least the scenario permits us to frame the discussion.

The literature contains discussion of other proposals for
stabilizing the moduli, which are alternatives to the racetrack
scheme. They all suffer, however, from difficulties similar to
those which have been discussed here.  We will not attempt a
complete review, but mention a few examples.

Ref.~\cite{nilles} focuses specifically on the heterotic string
dilaton, though similar arguments can be applied to other moduli.
It is argued that modifications of the gauge coupling function
along with gaugino condensation can lead to stabilization. SL(2,Z)
duality is used to significantly constrain the form of this
function, as well as the form of the \Kahler potential.  Not
surprisingly, however, this leads to stabilization at values of
the coupling ($\alpha$) of order one.  It is argued, there, that
uncomputable corrections might give a phenomenologically
acceptable value for the coupling.  However, this means precisely
that nothing is computable in such a picture.  The smallness of
the low energy gauge couplings is an accident; no aspect of string
(M-) theory dynamics is accessible to a systematic, weak coupling
analysis.

In ref.~\cite{dvali}, another low energy mechanism for stabilizing
the moduli is proposed. In this scheme, the low energy theory, in
a certain approximation, has a quantum-modified moduli space.  The
authors only consider the global limit; their models do not yield
supersymmetry-conserving minima with vanishing
cosmological constant when coupled to gravity.  Ignoring
this issue, as these authors note, the generic value of the gauge
coupling is of order one.  It is possible to obtain smaller
couplings, if the models have large discrete symmetries. These
models are not sufficiently developed to decide whether a weak
coupling analysis is applicable, but we would argue it is unlikely
that any sort of perturbative treatment of non-holomorphic
quantities is possible. If free of anomalies, large discrete
$R$-symmetries require large gauge groups (or large matter
content) . These lead to problems of calculability identical to
those we have described above.  Even if anomalies are cancelled by
a Green-Schwarz mechanism, it is difficult to avoid such large
groups.

\section{Conclusions}

Conceptually, the difficult issue in understanding how moduli are
stabilized in string theory is understanding why couplings are
weak and unified, and there are large hierarchies.  It is
certainly not hard to imagine that moduli are stabilized in such a
way that couplings and dimensionless ratios are of order one.  How
large pure numbers arise in a theory without small parameters is
distinctly more puzzling.  The racetrack scheme and its variants
which have been reviewed here are probably the most concrete
proposals for how moduli are stabilized at weak gauge coupling. We
have seen that in order that one obtain small couplings in a
controllable approximation, some degree of fine tuning is
required:  if gaugino condensation is the origin of the moduli
superpotential, one must have at least three gauge groups with
closely related $\beta$-functions; in theories with an unbroken
discrete $R$ symmetry, one needs two groups and an elaborate field
and symmetry structure.  In these cases, not everything is
calculable, but holomorphic quantities such as the superpotential
and the gauge coupling functions may be.    If supersymmetry is
broken at intermediate energy scales, many quantities will not be
computable.  If supersymmetry is broken at low energies, it is
likely that many quantities relevant to low energy physics
could be.

Comparing with other proposals for modulus stabilization in string
theory, the racetrack model has a certain appeal.  While the
fine-tuning is unattractive, and we don't have explicit examples
which provide a complete realization, the scenario is quite
concrete.  As we have described here, it offers the hope of
computing the gauge couplings and the superpotential.  If
supersymmetry is broken at low energies, many
quantities relevant to low energy physics might be computable.
\Kahler stabilization, by contrast, invokes uncontrollable and
unknown corrections to the \Kahler potential.
 As in the racetrack scenario, certain holomorphic quantities are calculable,
 but not the gauge couplings.  It is hard to reconcile this
 mechanism with low energy supersymmetry breaking.
 Maximally enhanced
symmetry, while possessing a certain economy, requires that
through some mysterious mechanism the gauge couplings are quite
small, and while it does suggest low energy supersymmetry
breaking, is not likely to offer the hope of computing even
holomorphic quantities.

There has been much interest recently in the possibility of large
internal dimensions.  The first well-developed proposal of this
type appeared in~\cite{wittencompact}, where it was assumed that
the strong coupling limit of the heterotic string, with
compactification on a Calabi-Yau space was appropriate.  The
difficulty with this idea, noted in~\cite{banksdinescales} is that
at large radius, the supersymmetry of the higher dimensional
theory insures that the potential vanishes.  Clearly, however,
something like the racetrack picture could operate here as well\cite{lalak}
In the scheme of \cite{wittencompact}, in particular, one has two walls,
and supersymmetry breaking arises from dynamics in the walls.
Two competing groups could give rise to a potential for the moduli,
with a minimum for some large value of some of the radii.  In this
regime, again, the \Kahler potential would not be calculable, but
the gauge couplings and other
holomorphic quantities would be.  This would presumably mean that
one could not take the geometric picture too literally;
at best, it would only be qualitatively correct.  Similar remarks
apply to other scenarios in which stabilization for large values
of geometrical moduli is required, e.g. \cite{rs}.

It is also interesting to consider the possibility that the
dilaton of this picture is an inflaton (these remarks are inspired by
the recent work of \cite{bankscosmology}
 as well as earlier work of \cite{bbs}).
In the models here, the dilaton is fixed, with a mass large compared
to the expected scale of supersymmetry breaking.  This is perhaps
promising, since for inflation one wants a rather large scale.
Moreover, the potential, as a result of the discrete fine tuning,
is rather flat.  Determining whether or not inflation occurs, and the
values of the relevant scales, requires an understanding of the
\Kahler potential, which we have argued is not calculable in these
schemes.  In any case,
the scale of the potential can quite naturally be
two or three orders of magnitude below the Planck scale,
perhaps
leading to a suitable spectrum of density fluctuations.
Determining the number of e-foldings also requires
knowledge of the \Kahler potential, but it is plausible
that this number might be large.  Because the potential
is a function of $S/N$, taking the \Kahler potential
to be of order $(N)^0$, a simple scaling argument
gives $N_e \sim N^2$.  In other words, the fine tuning required
to obtain a weak gauge coupling might be the same as the
fine tuning required to obtain adequate inflation.

Inflation in this picture requires explicit, if plausible, assumptions
about the \Kahler potential.  Still, the connection of the tuning
required to obtain small gauge couplings in this picture and that required to
obtain inflation raises the possibility that the two are truly
correlated;
perhaps the explanation of the smallness of the gauge couplings is
that
only regions with weak gauge couplings inflate.

\noindent {\bf Acknowledgements:}

\noindent
We thank T. Banks and N. Seiberg for comments and
a critical reading of an early version of the manuscript.
We are particularly grateful to I. Antoniadis for raising
the question of scalings in theories other than the
weakly coupled heterotic string.
The work of M.D. was supported in part by the U.S. Department of
Energy. The work of Y.S. was supported in part by NSF grant
PHY-9802484.

%%%%%%%%%%%%%%%%%%%%%%%%%%%%%%
%  Bibliography
%%%%%%%%%%%%%%%%%%%%%%%%%%%%%%

\end{document}